\documentclass[fleqn,usenatbib]{mnras}

\usepackage{newtxtext,newtxmath}

\usepackage[T1]{fontenc}

\DeclareRobustCommand{\VAN}[3]{#2}
\let\VANthebibliography\thebibliography
\def\thebibliography{\DeclareRobustCommand{\VAN}[3]{##3}\VANthebibliography}

\usepackage{graphicx}	
\usepackage{amsmath}	
\usepackage[utf8]{inputenc}

\newcommand{\software}[1]{\mbox{\sc {#1}}}

\title[BEBOP-3]{Fundamental effective temperature measurements for eclipsing binary stars -- VI. Improved methodology and application to the circumbinary planet host star BEBOP-3}

\author[P. F. L. Maxted et al.]{
P. F. L. Maxted,$^{1\,\href{https://orcid.org/0000-0003-3794-1317}{\includegraphics[scale=0.5]{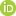}}}$
N. J. Miller,$^{2\,\href{https://orcid.org/0000-0001-9550-1198}{\includegraphics[scale=0.5]{orcid.jpg}}}$ 
T. A. Baycroft,$^{3\,\href{https://orcid.org/0000-0002-5510-8751}{\includegraphics[scale=0.5]{orcid.jpg}}}$
D. Sebastian,$^{4\,\href{https://orcid.org/0000-0002-2214-9258}{\includegraphics[scale=0.5]{orcid.jpg}}}$ 
A. H. M. J. Triaud$^{3\,\href{https://orcid.org/0000-0002-5510-8751}{\includegraphics[scale=0.5]{orcid.jpg}}}$
\and
and D.~V. Martin$^{5,6\,\href{https://orcid.org/0000-0002-7595-6360}{\includegraphics[scale=0.5]{orcid.jpg}}}$
 \\
$^{1}$Astrophysics Group, Keele University, Staffordshire ST5 5BG, UK\\
$^{2}$Observational Astrophysics, Department of Physics and Astronomy, Uppsala University, Box 516, 751 20, Uppsala, Sweden\\
$^{3}$School of Physics and Astronomy, University of Birmingham, Edgbaston, Birmingham B15 2TT, UK\\
${^4}$Th\"{u}ringer Landessternwarte Tautenburg, Sternwarte 5, D-07778, Tautenburg, Germany\\
$^{5}$Department of Physics \& Astronomy, Tufts University, Medford, MA 02155, USA\\
$^{6}$Department of Astronomy, The Ohio State University, Columbus, OH 43210, USA\\
}
\date{Accepted XXX. Received YYY; in original form ZZZ}

\pubyear{2024}

\begin{document}
\label{firstpage}
\pagerange{\pageref{firstpage}--\pageref{lastpage}}
\maketitle

\begin{abstract}
 BEBOP-3 is detached eclipsing binary star that shows total eclipses of a faint M~dwarf every 13.2 days by a 9$^{\rm th}$-magnitude F9\,V star. 
High precision radial velocity measurements have recently shown that this binary star is orbited by a  planet with an orbital period $\approx 550$~days. 
The extensive spectroscopy used to detect this circumbinary planet has also been used to directly measure the masses of the stars in the eclipsing binary.
We have used light curves from the TESS mission combined with these mass measurements to directly measure the following radii and surface gravities for the stars in this system:  $R_1 = 1.386  \pm 0.010\,R_{\odot}$,  $\log g_1 = 4.190  \pm 0.004$, $R_2 = 0.274  \pm 0.002\,R_{\odot}$, $\log g_2 = 4.979  \pm 0.002$.
We describe an improved version of our method to measure the effective temperatures (T$_{\rm eff}$) of stars in binary systems directly from their angular diameters and bolometric fluxes.
We measure T$_{\rm eff,1} = 6065{\rm\,K} \pm 44$\,K and T$_{\rm eff,2} = 3191{\rm\,K} \pm 40$\,K for the stars in BEBOP-3 using this method.
BEBOP-3 can be added to our growing sample of stars that can be used test the accuracy of spectroscopic and photometric methods to estimate  T$_{\rm eff}$ and $\log g$ for solar-type stars.
\end{abstract}

\begin{keywords}
techniques: spectroscopic, binaries: eclipsing, stars: fundamental parameters, stars: solar-type,stars: individual: BD+79 230
\end{keywords}



\section{Introduction}

\citet{2022ApJ...927...31T} have estimated the current level of realistic uncertainties on the fundamental parameter estimates for FGK main-sequence, subgiant, and lower giant branch stars that rely on empirical relations and stellar models.
They conclude that effective temperatures (T$_{\rm eff}$) estimates for these stars have a minimum uncertainty $\approx 2.4$\,per~cent. 
This uncertainty in T$_{\rm eff}$ and other sources of uncertainty, e.g. systematic errors in stellar models,  lead to uncertainties $\ga 5$\,per~cent in mass estimates for planet host stars and $\ga 20$\,per~cent in their ages.
This current level of uncertainty in the ages of planet host stars is at least twice the limit of $\pm 10$\,per~cent on the accuracy of in the ages of planet host stars that is a requirement of the European Space Agency PLATO mission \citep[PLAnetary Transits and Oscillations of stars,][]{2025ExA....59...26R}.
Both \citeauthor{2022ApJ...927...31T} and \citeauthor{2025ExA....59...26R} point to double-lined detached eclipsing binary stars as a promising source of fundamental data that can be used to improve this situation.
Accurate T$_{\rm eff}$ estimates are also essential to obtain accurate abundance measurements from the analysis of stellar spectra. 
\cite{2019ARA&A..57..571J}, in their review of the accuracy and precision of ``industrial scale'' stellar abundance measurements from large spectroscopic surveys, note that that it is not unusual to see differences of 200\,--\,300\,K in T$_{\rm eff}$ estimates for FGK-type stars obtained from different methods.

BEBOP-3 (BD+79~230) is an F9\,V star with an M-dwarf companion star that transits the primary star once every 13.2 days. 
The eclipses were found in ground-based photometry by the KELT survey \citep[Kilo-degree Extremely Little Telescope;][]{2007PASP..119..923P}. 
Radial velocity measurements showed that the transiting companion is too massive to be an exoplanet and so it was flagged as ``false-positive'' \citep{2018AJ....156..234C}. 
Many of these eclipsing binaries with a low-mass companion (``EBLM'' systems) have been identified by surveys for transiting exoplanets. 
Follow-up observations of these systems has become an active topic of research in its own right with implications for our understanding of the properties of very low mass stars (VLMSs), the formation and dynamical evolution of binary stars, circumbinary planets, etc. \citep{2023Univ....9..498M}. 

BEBOP-3 was selected for follow-up with extensive high-precision radial velocity (RV) measurements by the BEBOP project \citep[Binaries Escorted By Orbiting Planets;][]{2019A&A...624A..68M}, resulting in the discovery of a circumbinary planet with a mass of $\approx 0.56$\,$M_{\rm Jup}$ that orbits this eclipsing binary in an orbit with a period $P_{\rm b}\approx 550$\,d and an eccentricity of $e\approx0.25$ \citep{2025MNRAS.541.2801B}.
These RV measurements also hint at the presence of a second planet with an orbital period $\sim$1400\,d.
Although the M-dwarf companion contributes less than 0.5\,per~cent of the flux at optical wavelengths, \citeauthor{2025MNRAS.541.2801B} were able to measure the semi-amplitude of its spectroscopic orbit using over 120 spectra obtained with the SOPHIE spectrograph mounted on the 1.93-m telescope at the Observatoire de Haute-Provence \citep{2008SPIE.7014E..0JP}.
This measurement combined with the spectroscopic orbit of the F9\,V primary star leads directly to the following mass measurements for both stars: $M_1 = 1.083 \pm 0.026 M_{\odot}$;  $M_2 = 0.2615 \pm 0.0039\,M_{\odot}$. 

In this study, we combine the masses of the two stars measured by \cite{2025MNRAS.541.2801B} with an analysis of the TESS light curve to obtain precise and accurate radius measurements for the F9\,V primary and M-type secondary star in BEBOP-3. 
We then use the parallax of the system from Gaia DR3 \citep{2016A&A...595A...1G,2023A&A...674A...1G} combined with flux measurements at optical and infrared wavelengths to measure directly the effective temperature (T$_{\rm eff}$) of the F9\,V star. 
The use of Gaia DR3 parallaxes to directly measure T$_{\rm eff}$ for stars in eclipsing binary systems was first described and applied to the bright F9\,V\,+\,K0\,IV system AI~Phe by \citet{2020MNRAS.497.2899M}.
It has subsequently been applied to several other eclipsing binary systems, including the circumbinary planet system BEBOP-1 \citep[EBLM J0608$-$59;][]{2024MNRAS.531.4577M}.
Several improvements to our methodology for measuring  T$_{\rm eff}$ that have been used for this study of BEBOP-3 are described in Section~\ref{sec:teb}.
These direct, accurate and precise measurement of T$_{\rm eff}$ for a moderately bright stars ($V\approx11$) where the flux contribution from their M-dwarf companions is negligible makes these EBLM systems ideal benchmark stars that can be used to test the accuracy of T$_{\rm eff}$ and $\log g$  estimates for solar-type stars published by large spectroscopic surveys.

\section{Analysis}

\subsection{Mass and radius measurements}
The study of BEBOP-3 by \citeauthor{2025MNRAS.541.2801B} focusses on the discovery of a circumbinary planet and the direct mass measurements for the stars in this eclipsing binary system.  
They do not include radius measurements based on the stellar masses combined with the analysis of the light curve in their results because these were not needed for their study.
We do require accurate and precise radius estimates with robust error estimates for our analysis, so we describe that calculation here.

BEBOP-3 has been observed at 120-s cadence by TESS \citep{2015JATIS...1a4003R} in 10 sectors over 3 years. 
We have analysed this light curve using \software{jktebop}\footnote{\url{http://www.astro.keele.ac.uk/jkt/codes/jktebop.html}} \citep{2010MNRAS.408.1689S} assuming a power-2 limb-darkening law implemented in {\sc jktebop}  in terms of the parameters $h_1$ and $h_2$ \citep{2023Obs...143...71S}. The details of this analysis are very similar to those described in \citet{2024MNRAS.531.4577M}.

\begin{figure}
    \centering
    \includegraphics[width=1\linewidth]{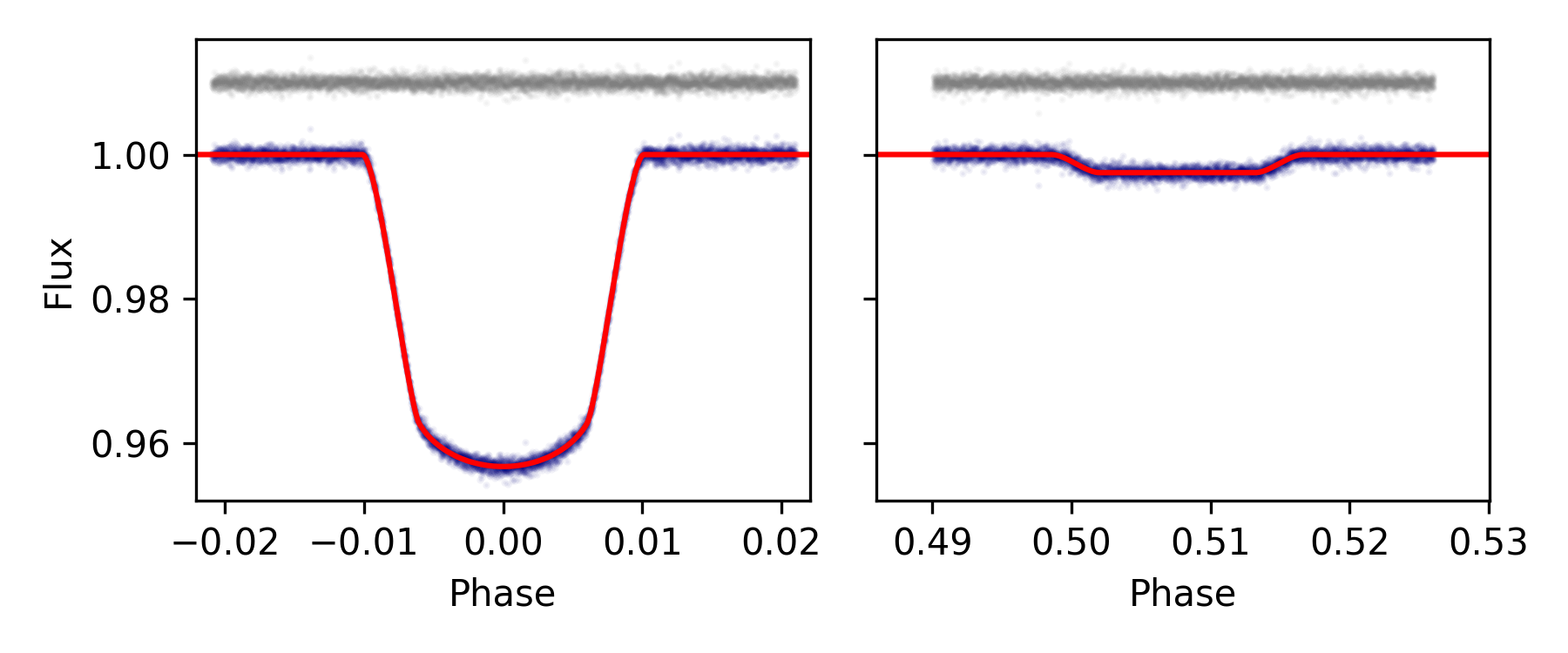}
    \caption{Photometry obtained from TESS images of BEBOP-3 as a function of orbital phase plotted with the best-fit light curve model. The residuals from the best-fit model are shown offset vertically above the light curve data.}
    \label{fig:lcfit}
\end{figure}

The value for the contamination of the photometric aperture provided in the meta data for the SPOC light curves is 0.3\,per~cent so we assume $\ell_3 = 0.003 \pm 0.001$ as a prior in the fit.
We also included priors on the values of $e\sin(\omega) = -0.06212\pm 0.0.00005$ and $e\cos(\omega)=0.01194 \pm 0.00006$ based on the spectroscopic orbit of the primary star from \citet{2025MNRAS.541.2801B}. 
The TESS light curve data and best-fit models for each subset are shown in Fig.~\ref{fig:lcfit} and the best-fit light curve parameters are given in Table~\ref{tab:lcfit}.

The best-fit parameters from these Monte Carlo simulations of the TESS light curve have then been combined with a random sample of  $K_1 = 19.364\pm0.001$\,km\,s$^{-1}$ and $K_2=80.22\pm0.74$\,km\,s$^{-1}$ values from a normal distribution based on the observed values with their standard errors from \cite{2025MNRAS.541.2801B} to infer the mass and radius of the stars with robust standard error estimates in nominal solar units \citep{2015arXiv151007674M} given in Table~\ref{tab:mrl}.

\begin{table*}
\caption[]{Light curve model parameters and  derived quantities from a least-squares fit to the TESS light curve of BEBOP-3. }
\label{tab:lcfit}
\begin{center}
\begin{tabular}{@{}lrl}
\hline
\noalign{\smallskip}
 \multicolumn{1}{@{}l}{Parameter} &
 \multicolumn{1}{l}{Value} &
 Note \\
\noalign{\smallskip}
\hline
\noalign{\smallskip}
BJD$_{\rm TDB}$ T$_0$ & $2459590.437745 \pm 0.000025 $ & Time of mid-transit \\
$P$    & $ 13.2176621 \pm 0.0000006 $ & Orbital period in days \\
$ (R_1+R_2)/a  $&$     0.06393 \pm     0.00009 $& Sum of the radii / semi-major axis of the binary orbit\\
$ k=R_2/R_1    $&$     0.19799 \pm     0.00015 $& \\
$ J_0          $&$     0.06981 \pm     0.00048 $& Surface brightness ratio at the centre of the stellar discs\\ 
$ h_1          $&$       0.802 \pm       0.002 $& Primary star limb-darkening parameter \\
$ i            $&$    88\fdg74 \pm     0\fdg01 $& Orbital inclination\\
$ \ell_3       $&$       0.003 \pm       0.001 $& Third light. Constrained by a prior.\\
$ e\cos(\omega)$&$    0.011946 \pm    0.000033 $& Constrained by a prior. \\
$ e\sin(\omega)$&$   -0.062121 \pm    0.000051 $& Constrained by a prior. \\
$ \ell_T       $&$     0.00253 \pm     0.00002 $& Flux ratio in TESS band\\
$ e            $&$    0.063259 \pm    0.000051 $& Orbital eccentricity \\
$ J            $&$     0.06449 \pm     0.00044 $& Average surface surface brightness ratio $=\ell_T/k^2$\\ 
$ r_1 = R_1/a  $&$    0.053370 \pm    0.000071 $& Fractional stellar radius, star 1\\
$ r_2 = R_2/a  $&$    0.010567 \pm    0.000019 $& Fractional stellar radius, star 2\\
\noalign{\smallskip}
\hline
\end{tabular}
\end{center}
\end{table*}

\begin{table}
\caption[]{Fundamental parameters of the stars in BEBOP-3. 
The metallicity [Fe/H] is taken from  \protect{\cite{2025MNRAS.541.2801B}. 
Note that the estimates of $\log(T_{\rm eff,2})$  and $\log(L_2)$ are strongly dependent on the use of semi-empirical colour\,--\,T$_{\rm eff}$ relations and so may subject to some additional systematic error -- see Section~\ref{sec:discuss}.}}

\label{tab:mrl}
\begin{center}
  \begin{tabular}{lrrr}
\hline
\noalign{\smallskip}
 \multicolumn{1}{l}{Parameter} &
 \multicolumn{1}{l}{Value} &
 \multicolumn{1}{l}{Error} &
 \multicolumn{1}{r}{} \\
\noalign{\smallskip}
\hline
\noalign{\smallskip}
$M_1/{\mathcal M^{\rm N}_{\odot}}$&1.084  & $\pm$ 0.026 & [2.4 \%] \\
\noalign{\smallskip}
$M_2/{\mathcal M^{\rm N}_{\odot}}$&0.262 & $\pm$ 0.004 &[1.5 \%] \\
\noalign{\smallskip}
$M_2/M_1$ & $ 0.2414 $ &$ \pm 0.0022$ &  [0.9 \%] \\
\noalign{\smallskip}
$R_1/{\mathcal R^{\rm N}_{\odot}}$&1.386 & $\pm$ 0.010 &[0.8 \%] \\
\noalign{\smallskip}
$R_2/{\mathcal R^{\rm N}_{\odot}}$&0.274 & $\pm$ 0.002 &[0.8 \%] \\
\noalign{\smallskip}
$\log(T_{\rm eff,1}/{\rm K}) $  & $ 3.7828$ & $\pm$ 0.0026  & [0.6 \%] \\
\noalign{\smallskip}
$\log(T_{\rm eff,2}/{\rm K)}$  & $ 3.5039 $ & $\pm$ 0.0044  & [1.0 \%] \\
\noalign{\smallskip}
$\rho_1/{\rho^{\rm N}_{\odot}}$  &0.407 & $\pm$ 0.002 &[0.5 \%] \\
\noalign{\smallskip}
$\rho_2/{\rho^{\rm N}_{\odot}}$  & 12.66 & $\pm$ 0.12 &[0.9 \%] \\
\noalign{\smallskip}
$\log g_1$ [cgs] & 4.1895 & $\pm$ 0.0041 & [1.0 \%]  \\
\noalign{\smallskip}
$\log g_2$ [cgs] & 4.9790 & $\pm$ 0.0016 & [0.4 \%] \\
\noalign{\smallskip}
$\log L_1/{\mathcal L^{\rm N}_{\odot}}$   & $0.3701 $&$ \pm 0.0061 $ &  [2.0\%]\\
\noalign{\smallskip}
$\log L_2/{\mathcal L^{\rm N}_{\odot}}$   & $-2.152 $&$ \pm 0.021$  & [4.8\%]\\
\noalign{\smallskip}
[Fe/H] & $-0.02$ & $\pm$ 0.10  & \\ 
\noalign{\smallskip}
\hline
\end{tabular}
\end{center}
\end{table}

\subsection{Reddening estimate}
To estimate the reddening towards BEBOP-3 we use the recently updated calibration of E(B$-$V) versus the equivalent width of the interstellar Na\,I~D lines by \citet{2025RNAAS...9..146M}.
This is an improved version of the calibration by \citet{1997A&A...318..269M}
To measure EW(Na\,I~D$_1$) and EW(Na\,I~D$_2$) we  applied the same method described in \citet{2024MNRAS.531.4577M} to 61 of the spectra obtain with the SOPHIE spectrograph observed through an airmass $<1.4$ and at times of low humidity. 
There are no interstellar absorption Na\,I lines visible in this high signal-to-noise spectrum with an equivalent width greater than about 10\,m\AA. 
Based on this analysis and the observed scatter in E(B$-$V) values for other stars with little on no interstellar Na\,I absorption from \citet{2025RNAAS...9..146M}, we decided use a Gaussian prior E(B$-$V)$=0 \pm 0.014$ in our analysis, excluding negative values of E(B$-$V).
A useful by-product of this analysis is a high signal-to-noise  spectrum of the primary star (S/N $\approx 900$ at 500\,nm).
This spectrum is available from the corresponding author's website\footnote{\url{https://www.astro.keele.ac.uk/pflm/BenchmarkDEBS/}} and in the supplementary online information that accompanies this article.

\subsection{Effective temperature measurements}
\label{sec:teb}

\subsubsection{Overview of our method}

The method we have developed to measure the effective temperature of stars in eclipsing binary systems is described in detail in  \citet{2020MNRAS.497.2899M}.
This method is not straightforward and will be unfamiliar to most readers, so we give here an overview that we hope will help the reader to understand our motivation for doing the calculation in this way.

The starting point for our calculation is the definition of effective temperature, ${\rm T}_{\rm eff}$. 
For a star with Rosseland radius $R$ and total luminosity $L$, ${\rm T}_{\rm eff}$ is defined by the equation  
\begin{equation}
\label{eq:teff}
{\rm T}_{\rm eff}  \equiv \left(L/(4\pi R^2 \sigma_{\rm SB})\right)^{1/4},
\end{equation}
where $\sigma_{\rm SB}$ is the Stefan-Boltzmann constant.\footnote {Note that this definition says nothing about the spectrum of the light emitted by the star, i.e. ${\rm T}_{\rm eff}$ is the effective temperature of a star whether or not it emits light as a black body.}
Our aim is to estimate ${\rm T}_{\rm eff,1}$ and ${\rm T}_{\rm eff,2}$, the effective temperature of the two stars in an eclipsing binary system, in a way that accounts for all known sources of uncertainty. 
One of the main uses for these direct measurements of ${\rm T}_{\rm eff}$ is to test the accuracy of methods based on stellar model atmospheres to estimate ${\rm T}_{\rm eff}$ so, to avoid a circular argument, we also want the results of our method to be insensitive to systematic errors in these models, e.g. missing opacity data leading to an underestimate of the line blanketing at ultraviolet wavelengths, the impact of magnetic fields on the temperature structure of the atmosphere, 3-dimensional effects due to turbulence and convection that are not captured accurately in 1-dimensional models, etc.

We use a Bayesian method to analyse the available data, $\bmath{D}$. The posterior probability distribution (PPD) to obtain these data given some model with parameters $\bmath{\Theta} = ({\rm T}_{\rm eff,1}, {\rm T}_{\rm eff,2}, \dots)$  is $P(\bmath{\Theta}| \bmath{D})\propto P(\bmath{D}|\bmath{\Theta})P(\bmath{\Theta})$, where $P(\bmath{\Theta})$ is the prior probability distribution for the model parameters, $\bmath{\Theta}$.
We use  \software{emcee} \citep{2013PASP..125..306F} to generate a large sample of points drawn from the PPD,  $P(\bmath{\Theta}| \bmath{D})$. 
We can then compute our best estimates for ${\rm T}_{\rm eff,1}$ and ${\rm T}_{\rm eff,2}$ from the mean values of these quantities in this sample with standard errors computed from the sample standard deviations. 

The data $\bmath{D}$ are the apparent magnitudes, photometric colour indices and  flux ratios measured at one or more wavelengths for a binary star system 
To calculate the the likelihood ${\cal L} =  P(\bmath{D}|\bmath{\Theta})$, we need to estimate the apparent spectral energy distributions (SEDs) of the two stars, ${\cal F}_{\lambda,1}$ and ${\cal F}_{\lambda,2}$,  so that we can integrate ${\cal F}_{\lambda,1} + {\cal F}_{\lambda,2}$ over the wavelength response function of the instruments used to measure the apparent magnitudes and  colours of the binary, and also integrate ${\cal F}_{\lambda,2}/{\cal F}_{\lambda,1}$ over the wavelength response function  of the instruments used for each observed flux ratio measurement. 
To calculate ${\cal F}_{\lambda,i}$ for star $i$ ($i=1,2$) we assume that the shape of the star's intrinsic (unreddened) SED is given by the normalised function $\tilde{f}_{\lambda,i}$, i.e., $\int_0^{\infty} \tilde{f}_{\lambda,i} d\lambda =1$. 
For a given estimate of $T_{\rm eff,i}$, equation (\ref{eq:teff}) leads to 
\begin{equation}
\label{eq:obsflux}
{\cal F}_{\lambda,i} = \frac{1}{4}\sigma_{\rm SB}\,\theta_i^2\,T_{\rm eff,i}^4\, \tilde{f}_{\lambda,i}\,A_{\lambda},
\end{equation}
where $A_{\lambda}$ is the extinction due to interstellar absorption as a function of wavelength and $\theta_i = 2R_i/d$ is the angular diameter of star $i$. 
The extinction $A_{\lambda}$ is calculated from an assumed value of the reddening, ${\rm E}({\rm B}-{\rm V})$, using a model for the extinction as a function of wavelength, so ${\rm E}({\rm B}-{\rm V})$ is one of the nuisance parameters in the vector of model parameters, $\bmath{\Theta}$. 
We include an estimate of ${\rm E}({\rm B}-{\rm V})$ in the prior, $P(\bmath{\Theta})$ because this improves the accuracy of the ${\rm T}_{\rm eff,1}$ and ${\rm T}_{\rm eff,2}$ measurements we obtain.
The angular diameters $\theta_1$ and $\theta_2$ are also included in $\bmath{\Theta}$ and their observed values with their covariance is included in the prior $P(\bmath{\Theta})$ so that we can account for the uncertainties in $R_1$, $R_2$ and the parallax $\varpi = 1/d$.

For the normalised functions that describe the shape of the intrinsic stellar SEDs, we assume $\tilde{f}_{\lambda,i} = N_i\,f^m_{\lambda,i}\Delta_{\lambda,i}$, where $f^m_{\lambda,i}$ is a model SED computed from a grid of stellar model atmospheres, $N_i$ is a constant defined such that $\int_0^{\infty} \tilde{f}_{\lambda,i} d\lambda =1$, and $\Delta_i$ is a smooth function that allows us to account for inaccuracies in the model SEDs. We use this approach so that $\tilde{f}_{\lambda,i}$ contain realistic stellar absorption features, e.g. Balmer lines, but their overall shape is determined by the observations. 
The polynomial coefficients $c_{i,1},c_{i,2},\dots$ that describe the smooth function $\Delta_i$ are also nuisance parameters that are included in the vector of model parameters, $\bmath{\Theta}$.

It is normally the case that we do not have direct measurements of the flux ratio across the full range of ultraviolet, optical and infrared wavelengths. 
Indeed, we only have direct measurement of the flux ratio in the TESS bandpass for BEBOP-3. 
This can lead the SEDs for the two stars being unrealistic if we do not put any constraints the ``distortion functions''  $\Delta_1$ and $\Delta_2$.
To constrain $\Delta_1$ and $\Delta_2$, we assume that the colours of the two stars in the binary are similar to the colours of other stars with the same effective temperature.
These ``flux ratio priors'' are computed by calculating the flux ratio in the Gaia G$_{\rm RP}$ band from ${\cal F}_{\lambda,2}/{\cal F}_{\lambda,1}$, and then using the using empirical colour\,--\,T$_{\rm eff}$ relations for the colours (${\rm NUV}-{\rm G}_{\rm RP}$), (${\rm G}_{\rm BP}-{\rm G}_{\rm RP}$), (G$_{\rm RP}-{\rm J}$), etc. applied to each star to infer the flux ratios at other wavelengths. These empirical flux ratio estimates with their standard error estimates can then be included in the calculation of $P(\bmath{\Theta}|\bmath{D})$.   
We chose the G$_{\rm RP}$ band as the reference wavelength for these colour\,--\,T$_{\rm eff}$ relations because it is very similar to the TESS photometric band, and we often have very precise flux ratio measurements for eclipsing binary stars based on the TESS light curve.

To calculate the likelihood, ${\cal L} =  P(\bmath{D}|\bmath{\Theta})$, we assume that the observed magnitudes, colours and flux ratios have independent Gaussian errors. 
We also account for the standard error in photometric zero-points that are used during our synthetic photometry calculations to convert fluxes to magnitudes or colours.
We find that we get typically get reduced chi-squared values $\chi^2_r>1$ for the fits to the observed data if we use the standard errors provided with the catalogue photometry and measured flux ratios.
This may be due to underestimated errors on the photometry, under-estimated errors on the photometric zero-points, inaccurate instrument response functions, intrinsic variability of the stars, etc. 
To account for this additional uncertainty, i.e., to achieve $\chi^2_r \approx 1$, we add additional variance $\sigma_m^2$ to the quoted variance of the observed magnitudes, and similarly for the variance of the observed colours ($\sigma_{\rm c}^2$)  and the variance of the observed flux ratios  ($\sigma_{\rm r}^2$).
These parameters of our noise model also are included in the $\Theta$.

\subsubsection{Updated methodology}
\label{sec:update}

The main improvement that we have made to our methodology is to compute a new set of  tabulated empirical colour\,--\,T$_{\rm eff}$ relations. 
These are described in Appendix~\ref{sec:color-teff}.
This avoids the need to compute a new set of polynomial colour\,--\,T$_{\rm eff}$ relation for each star, as we have done in previous studies. 
The standard deviation of the flux ratio prior at a given wavelength is set by the root-mean-square of the residuals for the stars used to calibrate the colour\,--\,T$_{\rm eff}$ relation.
This has been pre-computed and stored with the tabulated empirical colour\,--\,T$_{\rm eff}$ relations.

We have moved all the information needed to compute synthetic magnitudes out of the code itself and into a separate photometric database file. 
A new python script \software{ calspec.py} is now provided with the software that will download the filter response functions from the Spanish Virtual Observatory (SVO) filter profile service \citep{2024A&A...689A..93R}\footnote{\url{https://svo2.cab.inta-csic.es/theory/fps/}} and  spectra of flux standard stars from the CALSPEC database \citep{2025AJ....169...40B}.\footnote{\url{https://www.stsci.edu/hst/instrumentation/reference-data-for-calibration-and-tools/astronomical-catalogs/calspec}} 
These data are then used with observed magnitudes of the CALSPEC flux standard stars provided by the user to determine the photometric zero-points and other information needed to populate the photometric database file.
This makes it much easier to include photometry from a wide variety of sources.
We have found that moving all the information relevant to each star into a single input data file makes the software easier to use, particularly since the code itself can now be used to generate this file including photometry downloaded automatically from a variety of sources.
We have also added priors for  $\sigma_{\rm m}$,  $\sigma_{\rm c}$ and $\sigma_{\rm r}$ of the form
\[
P(\sigma_x) = \left\{
\begin{array}{ll}
\frac{1}{\alpha_x}e^{-\sigma_x/\alpha_x} & \sigma_x \ge 0, \\
0 & \sigma_x < 0, \\
\end{array}
\right.
\]
where the widths $\alpha_m$, etc.  are specified by the user in a configuration file. 
In previous applications of our method we used improper priors where we only required these hyper-parameters to be positive. 

The new version of the code also includes a calculation of the systematic errors in T$_{\rm eff}$, $\log L$ and de-reddened apparent flux ${\mathcal F}_{0,\lambda} = {\mathcal F}_{\lambda}/A_{\lambda}$ for each star due to the uncertainty in the absolute calibration of the CALSPEC flux scale \citep{2014PASP..126..711B}.
The details and results of the analysis are now stored in detail in a single output file that can be processed using \software{ Python} scripts provided with the software to produce diagnostic plots and output tables in \LaTeX\ format. 
The software is open-source and can be downloaded from \url{github.com/nmiller95/teb}. 
The results for BEBOP-3 presented here were computed with version v2025.08.26 of the software.
To test this updated version of our method, we repeated the analysis of 5 systems previous analysed using the same method implemented with older versions of our software. 
We re-calculated the prior on E(B$-$V) using the new calibrations for E(B$-$V) as a function of EW(Na~I D$_1$) and EW(D$_2$) from \citet{2025RNAAS...9..146M} for all stars except HD~22064.
These half-Gaussian priors have a standard deviation of about 0.014, which is larger than the standard deviations used in the original studies.
For HD~22064 we used the same prior on E(B$-$V)  based on Str\"{o}mgren photometry as \citet{2023MNRAS.522.2683M}. 
We did not include the FUV magnitude in our analysis for any of these stars because the observed values are all outside the magnitude range of the stars we used to calibrate the zero-point of the flux scale in this band. 
For AI Phe we used Gaia~DR3 photometry rather than the Gaia DR2 photometry used in the original study.
Otherwise, the observed magnitudes and flux ratios are the same as those used in the original studies. 
There is only one direct flux ratio measurement for HD~22064 and TOI-1338 so we set the width of the exponential prior on $\sigma_{\rm r}$ to $\alpha_{\rm r} = 0.005$. 
For the other stars we set this prior width to $\alpha_{\rm r} = 0.1$ so that value of  $\sigma_{\rm r}$ is mostly determined by the scatter of the measured flux ratios around the best fit model.
Similarly, we set the width of the prior for $\sigma_{\rm m}$ to $\alpha_{\rm m} = 0.1$. 
For AI~Phe, which is the only system for which photometric colours are included in the analysis, we set $\alpha_{\rm c} =0.1$ for the prior on $\sigma_{\rm c}$.

The results are summarised in Table~\ref{tab:teff_test}.
The agreement between the published results and our new methodology is excellent, and there is a noticeable improvement in the precision of the results for faint M-dwarf companions based on a single flux ratio measurements. 
Note that the results for these stars that have a single measurement of the flux ratio $\la 1$\,per~cent in the TESS band of are highly dependent on the assumed flux ratio priors computed from the assumed colour\,--\,T$_{\rm eff}$ relations because they emit most of their flux at near-infrared wavelengths.
The T$_{\rm eff}$ measurements for these stars may be subject to additional systematic error that is difficult to quantify.
We discuss this point further in Section~\ref{sec:discuss}.
The slight increase in T$_{\rm eff}$ for both stars in AI~Phe is mainly due to the less stringent prior on E(B$-$V).
This results in Table~\ref{tab:teff_test} demonstrate that our method is robust against the details of how it is implemented in software.

\begin{table}
\caption[]{Effective temperature measurements from published studies using our method compared to results from this work using our updated methodology. All values are in Kelvin. The standard error estimates given here are the sum of random and the systematic errors.}
\label{tab:teff_test}
\begin{center}
\begin{tabular}{@{}lrrl}
\hline
\noalign{\smallskip}
Star &
 \multicolumn{1}{l}{Published} & 
 \multicolumn{1}{l}{This work} &
 Notes \\
\hline
AI Phe A &$ 6199 \pm  34$  &$ 6259 \pm 42$ &  [Fe/H]$ \approx -0.14$ \\
AI Phe B &$ 5094 \pm  26$  &$ 5112 \pm 29$ & \citet{2020MNRAS.497.2899M}\\
\noalign{\smallskip}
CPD$-$54 810 A &$ 6462 \pm 56$  & $ 6462 \pm 49 $  &  [Fe/H]$ \approx 0.0$ \\
CPD$-$54 810 B &$ 6331 \pm 56$  & $ 6333 \pm 83 $  & \citet{2022MNRAS.517.5129M}\\
\noalign{\smallskip}
EBLM J0113+31 A & $ 6124 \pm 50$ &$ 6146 \pm 56 $ & [Fe/H]$ = -0.31$ \\
EBLM J0113+31 B & $ 3375 \pm 47$ &$ 3311 \pm 47 $ & \citet{2022MNRAS.513.6042M}\\
\noalign{\smallskip}
HD 22064 A & $ 6763 \pm  49 $ & $ 6711 \pm 62 $ &  [Fe/H]$ = -0.05$ \\
HD 22064 B & $ 3700 \pm 324 $ & $ 3486 \pm 79 $ & \citet{2023MNRAS.522.2683M}\\
\noalign{\smallskip}
TOI-1338 A & $ 6031 \pm  46 $ & $ 6054 \pm 68 $ &  [Fe/H]$ =  0.01$ \\
TOI-1338 B & $ 3220 \pm 135 $ & $ 3327 \pm 40 $ & \citet{2024MNRAS.531.4577M} \\
\noalign{\smallskip}
\hline
\end{tabular}
\end{center}
\end{table}

\subsubsection{Results for BEBOP-3}
The photometry used in this analysis is given in Table~\ref{tab:mags}. 
The Gaia photometry is from Gaia data release DR3 \citep{2016A&A...595A...1G, 2023A&A...674A...1G}. 
J, H and Ks magnitudes are from the 2MASS survey \citep{2006AJ....131.1163S}. 
WISE W1, W2 and W3 magnitudes \citep{2010AJ....140.1868W,2011ApJ...731...53M} are from the All-Sky Release Catalog.\footnote{\url{https://wise2.ipac.caltech.edu/docs/release/allwise/}}  
B$_{\rm T}$ and V$_{\rm T}$ magnitudes are taken from the Tycho-2 catalogue \citep{2000A&A...355L..27H}.
The parallax is taken from Gaia DR3 with corrections to the zero-point from \citet{2022MNRAS.509.4276F}.

We use model SEDs interpolated from a grid of synthetic spectra obtained from the Spanish Virtual Observatory\footnote{\url{http://svo2.cab.inta-csic.es/theory/newov2/index.php?models=bt-settl}} computed with BT-Settl model atmospheres \citep{2013MSAIS..24..128A} assuming  T$_{\rm eff}=6050$\,K, $\log g = 4.19$ for the primary star and T$_{\rm eff}=3100$\,K, $\log g = 4.98$ for the M-dwarf, and $[{\rm Fe/H}] = -0.02$ for both stars.
We found that using values of T$_{\rm eff}$ for the reference SED of the M~dwarf that differ by 100\,K or more from 3100\,K cannot reproduce the measured flux ratio in the TESS band, $\ell_{\rm T}$. 
We also noticed that the walkers in the \software{ emcee} sampler do not convergence to a well-defined solution if we use a reference SED for a T$_{\rm eff}$ value that does not produce a good fit to the observed value of $\ell_{\rm T}$.
We found that 6 distortion coefficients were needed to produce a model that gives a close match to the observed value of $\ell_{\rm T}$.
The value of T$_{\rm eff,1}$ varied by less than 10\,K between all these different models, i.e. the value of  T$_{\rm eff,1}$ that we measure is insensitive to the details of how we model the contribution of the M-dwarf companion.
This is not surprising given that the F9\,V star contributes more than 99.5\,per~cent of the total flux in this binary system.
In contrast, we only have one direct measurement of the properties of the M-dwarf (the flux ratio in the TESS band) and the results for this faint companion are strongly dependence on the flux ratio priors and reference SED used. 
For that reason, we caution that the values of T$_{\rm eff,2}$ and $\log L_2$ we have derived here may be subject to additional systematic error. 
We discuss this point further in Section~\ref{sec:discuss}.

\begin{figure}
\includegraphics[width=\columnwidth]{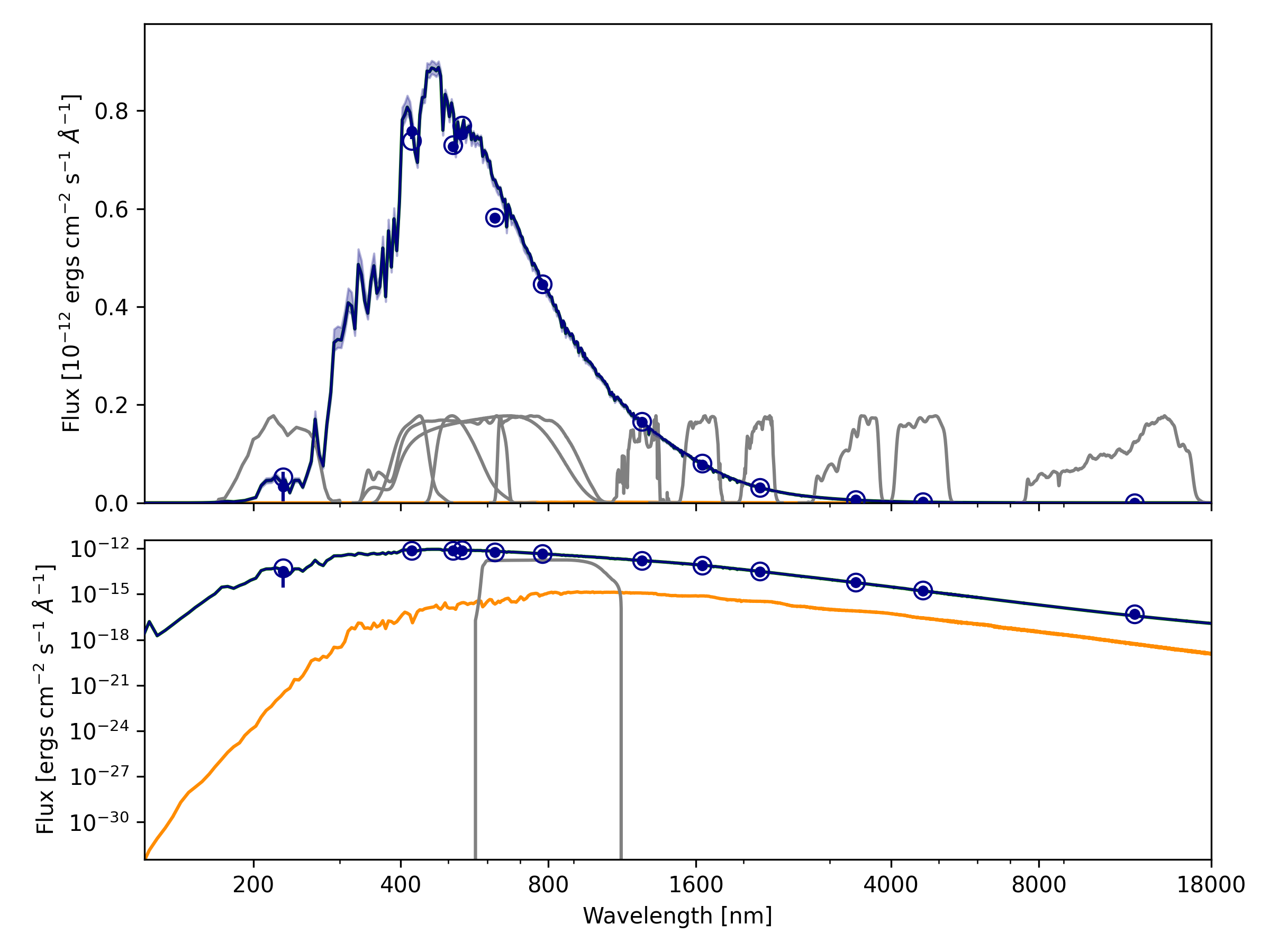}
\caption{Upper panel: The SED of BEBOP-3. The best-fit SED is plotted as a line and the
mean SED $\pm 1-\sigma$ is plotted as a filled region. The observed fluxes are
plotted as points with error bars and predicted fluxes for the best-fit SED
integrated over the response functions shown are plotted with open circles.
The SEDs of the two stars are also plotted but are barely distinguishable at this scale.
Lower panel: Same as the upper panel but with fluxes plotted on a logarithmic scale. 
The response function of the TESS instrument is also plotted here.}
\label{fig:sed}
\end{figure}

The predicted apparent magnitudes are compared to the observed apparent magnitudes in Table~\ref{tab:mags} and are shown in Fig.~\ref{fig:sed}.
The systemic errors given in Table~\ref{tab:mags} account for the uncertainty in the CALSPEC flux scale using the data provided in the file {\tt WDcovar\_002.fits} from the CALSPEC web site \citep{2014PASP..126..711B}.

\begin{figure}
    \centering
    \includegraphics[width=1\linewidth]{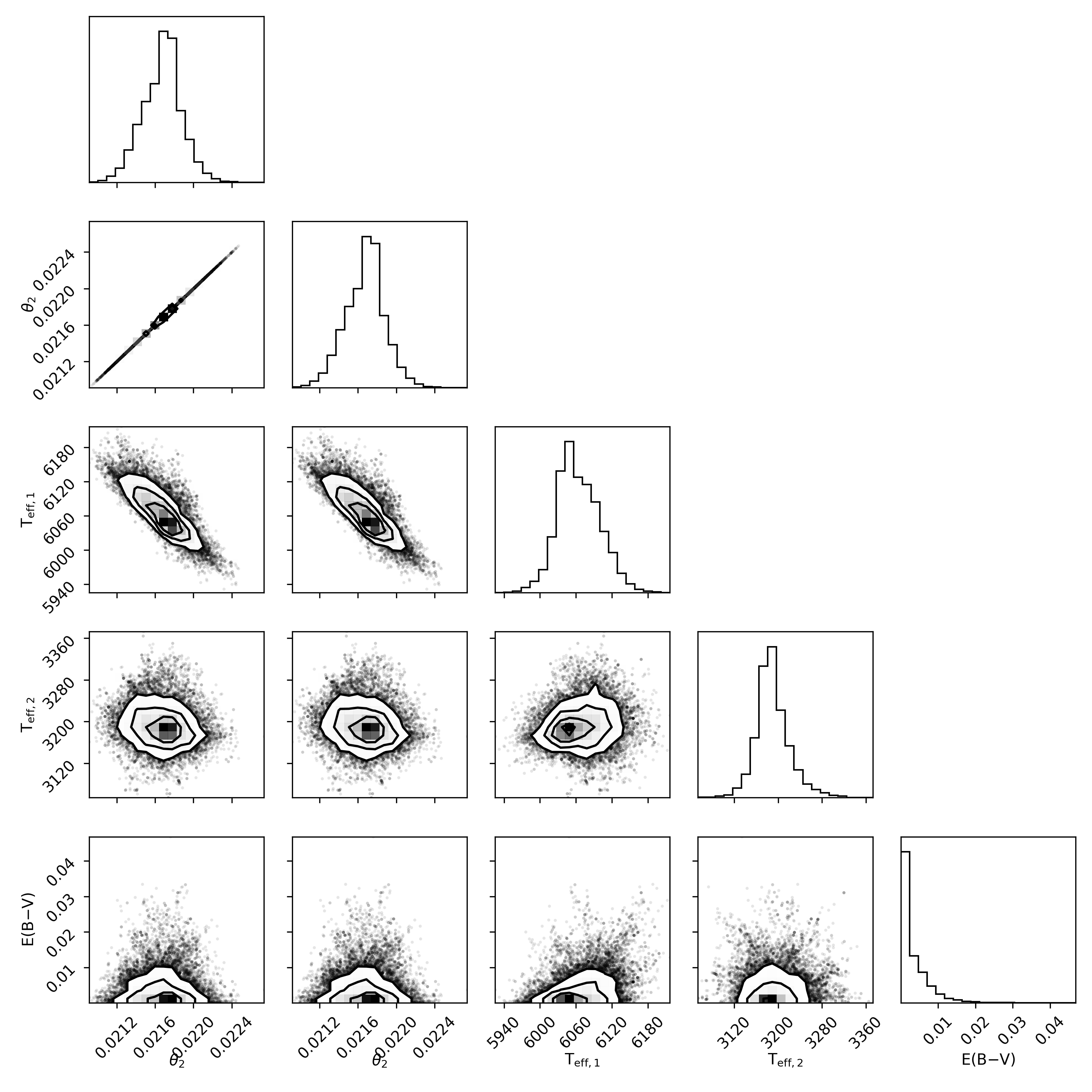}
    \caption{Pair-wise joint distributions and sample histograms for some parameters of interest from our analysis of BEBOP-3 plotted using \software{corner} \citep{corner}. }
    \label{fig:corner}
\end{figure}

The posterior probability distribution for the model parameters from our analysis to measure the stellar effective temperatures is summarised in Table~\ref{tab:teb} and the spectral energy distribution is plotted in Fig.~\ref{fig:sed}. 
The pair-wise joint distributions and sample histograms for some parameters of interest are shown in Fig.~\ref{fig:corner}. 
It can be seen that the distributions of T$_{\rm eff,1}$  and T$_{\rm eff,2}$ are slightly asymmetric. 
If we take the 15.9-percentile and 84.1-percentile points around the median values as asymmetric error estimates we obtain  T$_{\rm eff,1} = 6061^{+40}_{-29}$\,K  and T$_{\rm eff,2} = 3187^{+32}_{-22}$\,K. This asymmetry arises from the dependence of these T$_{\rm eff}$ estimates on the reddening estimate, E$(\rm B-\rm V)$, which naturally has an asymmetric distribution because this quantity cannot be negative.   
T$_{\rm eff,1}$  is more sensitive to the assumed value of E$(\rm B-\rm V)$ than T$_{\rm eff,2}$ because more of the primary star's flux is emitted at blue wavelengths.  
The full output from our analysis is available available in the supplementary online information that accompanies this article.

\begin{table*}
\caption{Observed magnitudes, colours and flux ratios for BEBOP-3 and predicted values based on our synthetic photometry. The
predicted magnitudes are shown with error estimates from the uncertainty on the zero-points for each photometric system.  The
pivot wavelength for each band pass is shown in the column headed  $\lambda_{\rm pivot}$.  The estimated apparent magnitudes for
each star are shown given in the columns headed $m_1$ and headed $m_2$.  The flux ratio in each band is shown in the final
column. }
\label{tab:mags}
\centering
\begin{tabular}{@{}lrrrrrrr}
\hline
Band &  $\lambda_{\rm pivot}$ [nm]& \multicolumn{1}{c}{Observed} &\multicolumn{1}{c}{Computed} &
\multicolumn{1}{c}{$\rm O-\rm C$} &\multicolumn{1}{c}{$m_1$}  & \multicolumn{1}{c}{$m_2$}  &
\multicolumn{1}{c}{$\ell$} [\%] \\
\hline
\noalign{\smallskip}
NUV &   230.1 & $14.560\pm 1.000 $& $14.050\pm 0.094 $& $+0.510 \pm 1.004 $& $14.050\pm 0.094 $ & $30.607\pm 0.094 $ &  0.00 \\
B$_{\rm T}$  &   421.2 & $ 9.895\pm 0.023 $& $ 9.923\pm 0.014 $& $-0.028 \pm 0.027 $& $ 9.923\pm 0.014 $ & $20.108\pm 0.014 $ &  0.01 \\
G$_{\rm BP}$ &   511.0 & $ 9.392\pm 0.003 $& $ 9.387\pm 0.006 $& $+0.004 \pm 0.007 $& $ 9.388\pm 0.006 $ & $18.215\pm 0.006 $ &  0.03 \\
V$_{\rm T}$  &   533.5 & $ 9.320\pm 0.017 $& $ 9.294\pm 0.014 $& $+0.026 \pm 0.022 $& $ 9.294\pm 0.014 $ & $18.174\pm 0.014 $ &  0.03 \\
G   &   621.8 & $ 9.108\pm 0.003 $& $ 9.106\pm 0.008 $& $+0.002 \pm 0.008 $& $ 9.107\pm 0.008 $ & $16.566\pm 0.008 $ &  0.10 \\
G$_{\rm RP}$ &   776.9 & $ 8.656\pm 0.004 $& $ 8.655\pm 0.004 $& $+0.002 \pm 0.006 $& $ 8.657\pm 0.004 $ & $15.333\pm 0.004 $ &  0.21 \\
J   &  1240.6 & $ 8.157\pm 0.023 $& $ 8.150\pm 0.015 $& $+0.007 \pm 0.028 $& $ 8.158\pm 0.015 $ & $13.550\pm 0.015 $ &  0.70 \\
H   &  1649.0 & $ 7.934\pm 0.040 $& $ 7.897\pm 0.019 $& $+0.037 \pm 0.044 $& $ 7.906\pm 0.019 $ & $13.032\pm 0.019 $ &  0.89 \\
Ks  &  2162.9 & $ 7.860\pm 0.017 $& $ 7.827\pm 0.030 $& $+0.033 \pm 0.035 $& $ 7.839\pm 0.030 $ & $12.771\pm 0.030 $ &  1.06 \\
W1  &  3389.7 & $ 7.805\pm 0.028 $& $ 7.783\pm 0.036 $& $+0.022 \pm 0.046 $& $ 7.798\pm 0.036 $ & $12.496\pm 0.036 $ &  1.32 \\
W2  &  4640.6 & $ 7.840\pm 0.019 $& $ 7.807\pm 0.059 $& $+0.033 \pm 0.062 $& $ 7.825\pm 0.059 $ & $12.269\pm 0.059 $ &  1.67 \\
W3  & 12567.5 & $ 7.814\pm 0.017 $& $ 7.787\pm 0.053 $& $+0.027 \pm 0.056 $& $ 7.803\pm 0.053 $ & $12.333\pm 0.053 $ &  1.54 \\
\noalign{\smallskip}
\multicolumn{5}{@{}l}{Flux ratios [\%]} \\
\noalign{\smallskip}
TESS &   788.0 & $ 0.253 \pm 0.002 $& $ 0.250 $& $+0.003 \pm 0.002 $ \\
\noalign{\smallskip}
\hline
\end{tabular}
\end{table*}


\begin{table}
\caption{Results from our analysis to measure the effective temperatures for both stars in
BEBOP-3. The output parameter values are calculated using the mean and standard error of the
posterior probability distribution sampled using {\tt emcee}. Note that the results for stars 2
are very dependent on the flux ratio priors calculated from our assumed colour\,--\,T$_{\rm
eff}$ relations, and so may be subject to additional systematic error.}
\label{tab:teb}
\centering
\begin{tabular}{@{}lr}
\hline
Parameter & \multicolumn{1}{l}{Value} \\
\hline
\noalign{\smallskip}
\multicolumn{2}{@{}l}{Priors} \\
$\theta_1$ [mas] & $0.10940 \pm 0.00085$ \\
$\theta_2$ [mas] & $0.02166 \pm 0.00017$ \\
E(B$-$V) & $0.000 \pm 0.014 $ \\
$\sigma_{\rm m}$& 0.1 \\
$\sigma_{\rm r}$& 0.005 \\
\noalign{\smallskip}
\multicolumn{2}{@{}l}{Model parameters} \\
$T_{\rm eff,1}$ [K] & $6065 \pm 35$ (rnd) $\pm 9$ (sys) \\
$T_{\rm eff,2}$ [K] & $3191 \pm 31$ (rnd) $\pm 9$ (sys) \\
$\theta_1$ [mas] & $0.1095 \pm 0.0010 $ \\
$\theta_2$ [mas] & $0.02167 \pm 0.00020 $ \\
E(B$-$V) &$ 0.003 \pm 0.004 $ \\
$\sigma_{\rm m} $&$ 0.010 \pm 0.012 $ \\
$\sigma_{\rm r} $&$ 0.002 \pm 0.003 $ \\
$c_{1,1} $&$ -0.001 \pm 0.129 $ \\
$c_{2,1} $&$  0.011 \pm 0.140 $ \\
$c_{1,2} $&$  0.116 \pm 0.193 $ \\
$c_{2,2} $&$ -0.029 \pm 0.164 $ \\
$c_{1,3} $&$  0.049 \pm 0.176 $ \\
$c_{2,3} $&$  0.080 \pm 0.164 $ \\
$c_{1,4} $&$ -0.042 \pm 0.142 $ \\
$c_{2,4} $&$  0.107 \pm 0.165 $ \\
$c_{1,5} $&$  0.097 \pm 0.094 $ \\
$c_{2,5} $&$ -0.007 \pm 0.130 $ \\
$c_{1,6} $&$ -0.080 \pm 0.112 $ \\
$c_{2,6} $&$  0.037 \pm 0.121 $ \\
\noalign{\smallskip}
\multicolumn{2}{@{}l}{Derived parameters} \\
${\mathcal F}_{\oplus,1}$[$10^{-8}$ erg\,cm$^{-2}$\,s$^{-1}$] & $0.5402 \pm 0.0072$ (rnd) $\pm 0.0033$ (sys)  \\
${\mathcal F}_{\oplus,2}$[$10^{-10}$ erg\,cm$^{-2}$\,s$^{-1}$] & $0.1623 \pm 0.0067$ (rnd) $\pm 0.0009$ (sys)  \\
$\log(L_1/L_{\odot})$  & $0.3701 \pm 0.0061$ (rnd) $\pm 0.0026$ (sys)  \\
$\log(L_2/L_{\odot})$  & $-2.152 \pm 0.018$ (rnd) $\pm 0.003$ (sys)  \\
\noalign{\smallskip}
\hline
\end{tabular}
\end{table}

\section{Discussion}
\label{sec:discuss}

\begin{figure}
\centering
\includegraphics[width=\linewidth]{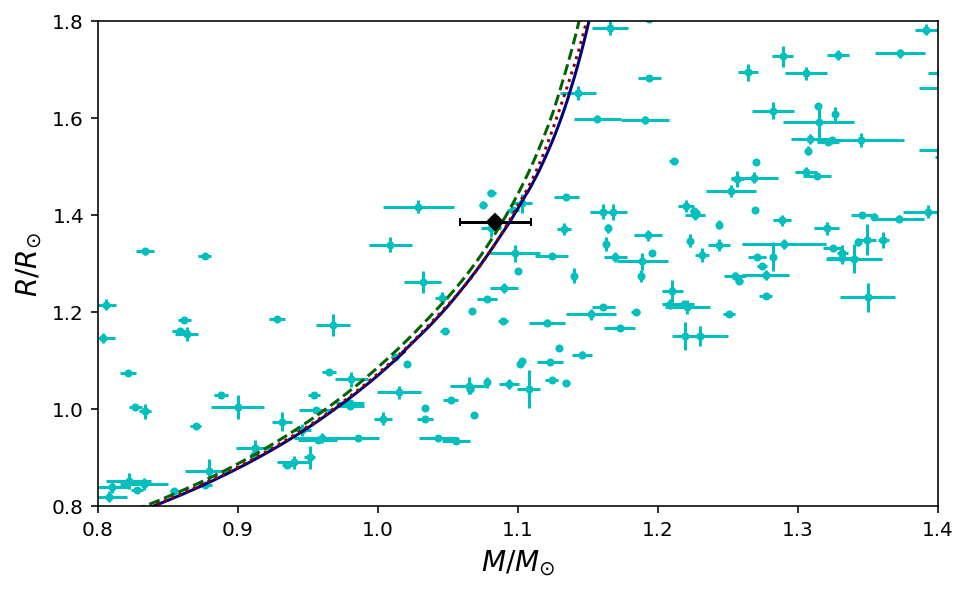}
\includegraphics[width=\linewidth]{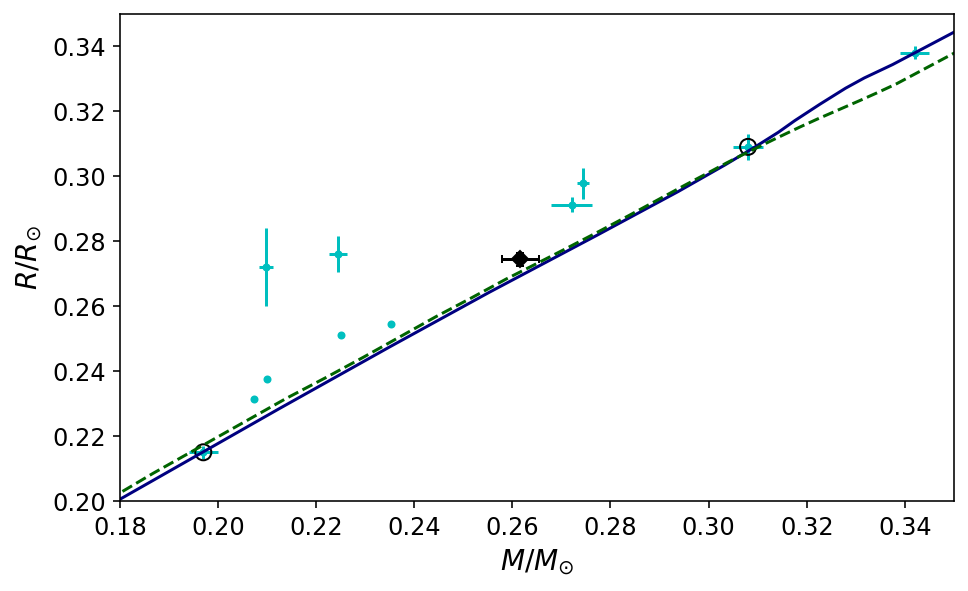}
\includegraphics[width=\linewidth]{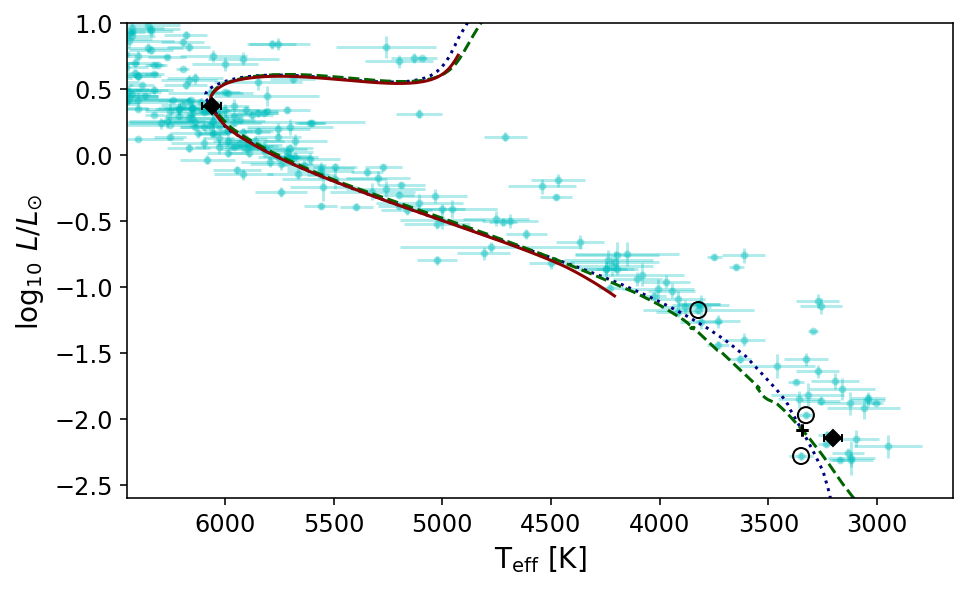}
\caption{ Upper panel: primary component of BEBOP-3 in the mass\,--\,radius plane (diamond with error bars). 
Middle panel: secondary component of BEBOP-3 in the mass\,--\,radius plane (diamond with error bars). 
Lower panel: both components of BEBOP-3 in the  Hertzsprung-Russell diagram (diamonds with error bars).  
The points on the isochrones corresponding to the observed mass of BEBOP-3\,B are marked with ``+'' symbols.
All panels show isochrones for an age of 5.9\,Gyr assuming $[{\rm Fe/H}]=0.0$ from \software{bagemass} (red solid line), DSEP (blue dotted line) and MIST (green dashed line). 
Cyan error bars show stars in eclipsing binary systems taken from DEBCat \citep{2015ASPC..496..164S}.
The values of T$_{\rm eff}$ and $\log L/L_{\odot}$ for the three M-dwarf companions in eclipsing binaries in Table~\ref{tab:teff_test} have been used here.
These stars are circled in the middle and lower panels.
}
\label{fig:hrd}
\end{figure}

Fig.~\ref{fig:hrd} shows the properties of the stars in BEBOP-3 compared to other stars in detached eclipsing binary systems and a selection of isochrones from stellar models. 
To estimate the age of BEBOP-3 we analysed the properties of primary star using \software{bagemass} \citep{2015A&A...575A..36M}. 
The details of this analysis are the same as described in \citet{2024MNRAS.531.4577M}.
Isochrones for the same age and initial metal abundance from the Dartmouth stellar evolution database \citep[DSEP, ][]{2008ApJS..178...89D} and the MESA Isochrones \& Stellar Tracks \citep[MIST, ][]{2016ApJ...823..102C} are also shown in Fig.~\ref{fig:hrd}.

Fig.~\ref{fig:hrd} shows the long-standing ``radius inflation'' problem, whereby the radii of low-mass stars tend to be under-predicted by stellar models \citep{2013ApJ...776...87S}. 
This is not the case for BEBOP-3\,B or the other M-dwarfs from Table~1 that are highlighted in Fig.~\ref{fig:hrd}.
These M-dwarfs have the advantage of being in long-period binary systems, unlike the majority of M-dwarf stars with accurate mass and radius measurements which tend to be found in binaries with an orbital period of a few days or less. 
These stars in short-period binary systems are forced by tidal interactions with their companions to rotate synchronously with their orbits, i.e. more rapidly than is typical for single M-dwarfs and M-type planet host stars \citep{2013MNRAS.432.1203M, 2023AJ....165..129B,2024AJ....168...93J}.
These results are broadly consistent with the idea that the radius inflation problem is related to the generation of magnetic fields by rotation in low mass stars \citep{2001ApJ...559..353M,2014ApJ...789...53F,2017AJ....153..101S}.

Although the three very low mass stars highlighted in Fig.~\ref{fig:hrd}, including BEBOP-3\,B, show good agreement with the stellar models shown, there are offsets of about 100\,K that appear to be significant if we do not take account of the additional systematic error that may be present due to the strong dependence of these T$_{\rm eff,2}$ measurements on the assumed colour\,--\,T$_{\rm eff}$ relations.
This offsets may be related to the difference in  [Fe/H] for these stars estimated from the analysis of the primary star spectrum in these systems (Table~\ref{tab:teb}).
However, these [Fe/H] estimates have been measured using a variety of different techniques and so are subject to systematic errors.
To properly test the accuracy of isochrones for M dwarfs, we need accurate T$_{\rm eff,2}$ measurements based on observations of the eclipse depth at near-infrared wavelengths, i.e. around the peak of the M-dwarf's SED, combined with consistent estimates for [Fe/H] from the host star spectrum.

\section{Conclusion}
BEBOP-3 adds to a small but growing sample of eclipsing binaries for which we have directly measured the effective temperature (T$_{\rm eff}$) and surface gravity ($\log g$) of a solar-type star with a much fainter M-dwarf companion. 
These systems can be used to test the accuracy of methods used to estimate T$_{\rm eff}$ and $\log g$ using spectroscopic and/or photometric methods, e.g. the results provided in catalogues produced by large spectroscopic surveys.
The software we have used to implement our method has been updated so that it is now much easier to use.
We find that our effective temperature measurements for stars in eclipsing binaries we have made using this method are not sensitive to the details of the way that our method is implemented in software, i.e. the results we obtain here are consistent with our analysis
using the previous versions of our software shown in Table~\ref{tab:teff_test} despite the
substantial changes to the software described in Section~\ref{sec:update}.

\section*{Acknowledgements}

We thank the anonymous referee for their comments that have improved the manuscript.

We are grateful to Luca Casagrande for providing us with the data from the GALAH survey that we have used in Appendix~\ref{sec:color-teff}

This research is supported work funded from the European Research
Council (ERC) under the European Union’s Horizon 2020 research
and innovation programme (grant agreement No. 803193 - BEBOP).

PM acknowledges support from UK Science and Technology Facilities Council (STFC) research grant numbers ST/Y002563/1 and UKRI1193.

This work has made use of data from the European Space Agency (ESA) mission
{\it Gaia} (\url{https://www.cosmos.esa.int/gaia}), processed by the {\it Gaia}
Data Processing and Analysis Consortium (DPAC,
\url{https://www.cosmos.esa.int/web/gaia/dpac/consortium}). Funding for the DPAC
has been provided by national institutions, in particular the institutions
participating in the {\it Gaia} Multilateral Agreement.

This paper includes data obtained through the TESS Guest investigator programs  G06022 (PI Martin), G05024 (PI Martin), G04157 (PI Martin), G03216 (PI Martin) and G022253 (PI Martin).

This paper includes data collected by the TESS mission, which is publicly available from the Mikulski Archive for Space Telescopes (MAST) at the Space Telescope Science Institure (STScI). Funding for the TESS mission is provided by the NASA Explorer Program directorate. STScI is operated by the Association of Universities for Research in Astronomy, Inc., under NASA contract NAS 5–26555. We acknowledge the use of public TESS Alert data from pipelines at the TESS Science Office and at the TESS Science Processing Operations Center.

This research made use of Lightkurve, a Python package for Kepler and TESS data analysis \citep{2018ascl.soft12013L}.

This research has made use of the SIMBAD database, the VizieR catalogue access tool, and the cross-match service provided by CDS, Strasbourg, France. 

This publication makes use of data products from the Wide-field Infrared Survey Explorer, which is a joint project of the University of California, Los Angeles, and the Jet Propulsion Laboratory/California Institute of Technology, and NEOWISE, which is a project of the Jet Propulsion Laboratory/California Institute of Technology. WISE and NEOWISE are funded by the National Aeronautics and Space Administration.
 
This publication makes use of data products from the Two Micron All Sky Survey, which is a joint project of the University of Massachusetts and the Infrared Processing and Analysis Center/California Institute of Technology, funded by the National Aeronautics and Space Administration and the National Science Foundation.

Based on data retrieved from the SOPHIE archive at Observatoire de Haute-Provence (OHP), available at \url{atlas.obs-hp.fr/sophie}.

\section*{Data Availability}
The data underlying this article are available in the following repositories:  
Mikulski Archive for Space Telescopes -- \url{https://archive.stsci.edu/}  (TESS); SOPHIE archive at Observatoire de Haute-Provence (OHP) -- \url{atlas.obs-hp.fr/sophie}; VizieR catalogue access tool, CDS, Strasbourg, France -- \url{https://vizier.cds.unistra.fr/}.

\appendix
\section{New empirical colour -- effective temperature relations}
\label{sec:color-teff}

We have used two samples of stars selected from the GALAH survey \citep[Galactic Archaeology with HERMES;][]{2021MNRAS.506..150B} and the CARMENES survey  \citep[Calar Alto high-Resolution search for M dwarfs with Exoearths;][]{2014SPIE.9147E..1FQ} to calibrate empirical colour -- effective temperature relations for A- to K-type stars and M-type stars, respectively. 
The sample of stars observed by the GALAH survey is the same as that described in \citet{2021MNRAS.507.2684C} and the sample of M-dwarf stars observed by the CARMENES survey is the same as that described in \citet{2020A&A...642A.115C}. 
For the GALAH survey, we applied to the following selection criteria to obtain a sample of dwarf stars with low reddening and with reliable measurements:
\begin{itemize}
\item ${\tt ebv}<0.1$;
\item ${\tt galah\_flag\_sp = 0}$;
\item ${\tt dr3\_phot\_proc\_mode=0}$;
\item $-0.08 < {\tt dr3\_phot\_bp\_rp\_excess\_factor\_corr} < 0.2$;
\item ${\tt galah\_logg} > 3.5$.
\end{itemize}
The exact definitions of these quantities are given in \citet{2021MNRAS.507.2684C}.

We used the Vizier ``X-Match'' service\footnote{\url{http://cdsxmatch.u-strasbg.fr/}} to obtain photometric measurements at ultraviolet and near-infrared wavelengths for the stars in the two samples the following surveys: GALEX \citep[FUV, NUV;][]{2017ApJS..230...24B}; 2MASS \citep[J, H, Ks;][]{2011ApJ...735..112J}; ALLWISE  \citep[W1, W2, W3;][]{2010AJ....140.1868W,2011ApJ...731...53M}. 
For the GALAH sample, we used the de-reddened photometry in the Gaia G$_{\rm BP}$ and G$_{\rm RP}$ bands included with the catalogue from \citet{2021MNRAS.507.2684C}. 
For the CARMENES sample, we used the Vizier ``X-Match'' service to obtain photometry for each star in the Gaia G$_{\rm BP}$ and G$_{\rm RP}$ bands from the Gaia DR3 catalogue \citep{2016A&A...595A...1G, 2023A&A...674A...1G}.
We assume that the reddening for these nearby M-dwarfs is negligible.

We removed blended sources from the samples used to compute the colour\,--\,T$_{\rm eff}$ relations in the ALLWISE W1, W2 and W3 bands  by excluding stars that have more than one source listed in the 2MASS catalogue within 12.2\arcsec, 12.8\arcsec\ or 13.0\arcsec, respectively, of the ALLWISE source position.
These exclusion limits were selected because they are equal to the resolution of the instrument at these wavelengths.\footnote{\url{https://wise2.ipac.caltech.edu/docs/release/prelim/}}
We did not include the ALLWISE W4 band in these empirical colour\,--\,T$_{\rm eff}$ relations because there is a clear discontinuity between the relations defined by the GALAH and CARMENES samples. 
This may be because of the large photometric apertures used to measure the flux in the W4 band (up to 49.5\arcsec) or contamination of the photometry by dust emission at this wavelength.\footnote{\url{https://wise2.ipac.caltech.edu/docs/release/allsky/expsup/sec1_4b.html}} 

The distribution of stars in the $({\rm FUV}-{\rm NUV})$ --  $({\rm G}_{\rm BP} - {\rm G}_{\rm RP})$ colour--colour diagram shows a ``tail'' of stars that are red at optical wavelengths and blue at ultraviolet wavelengths. 
We assume that these are blended sources, e.g. binary systems with a hot white dwarf or hot subdwarf component.
We exclude these stars from the sample used to compute the colour\,--\,T$_{\rm eff}$ relations in the GALEX NUV and FUV bands by rejecting stars  with $({\rm FUV}-{\rm NUV}) <  6.5 \times ({\rm G}_{\rm BP} - {\rm G}_{\rm RP})$

The effective temperatures estimates for the M-dwarfs in the CARMENES sample are quoted to the nearest 100\,K, so we use sub-samples of stars within each 100\,K bin to compute a mean colour and standard error estimate in 27 bins from 2500\,K to 4700\,K. 
Outlier values more than 5 times the median absolute deviation from the median value in each bin were excluded from the calculation of the mean and sample standard deviation in each bin for each colour. For the GALAH sample, we used used the effective temperature estimated using the infrared flux method ({\tt Teff\_irfm}) to assign each star to one of 53 bins of width 50\,K from 3800\,K to 9000\,K. There are more data per T$_{\rm eff}$ bin for this sample so outlier values more than 10 times the median absolute deviation from the median value in each bin were excluded from the calculation of the mean and sample standard deviation in each bin for each colour. 

\begin{figure*}
	\includegraphics[width=0.75\textwidth]{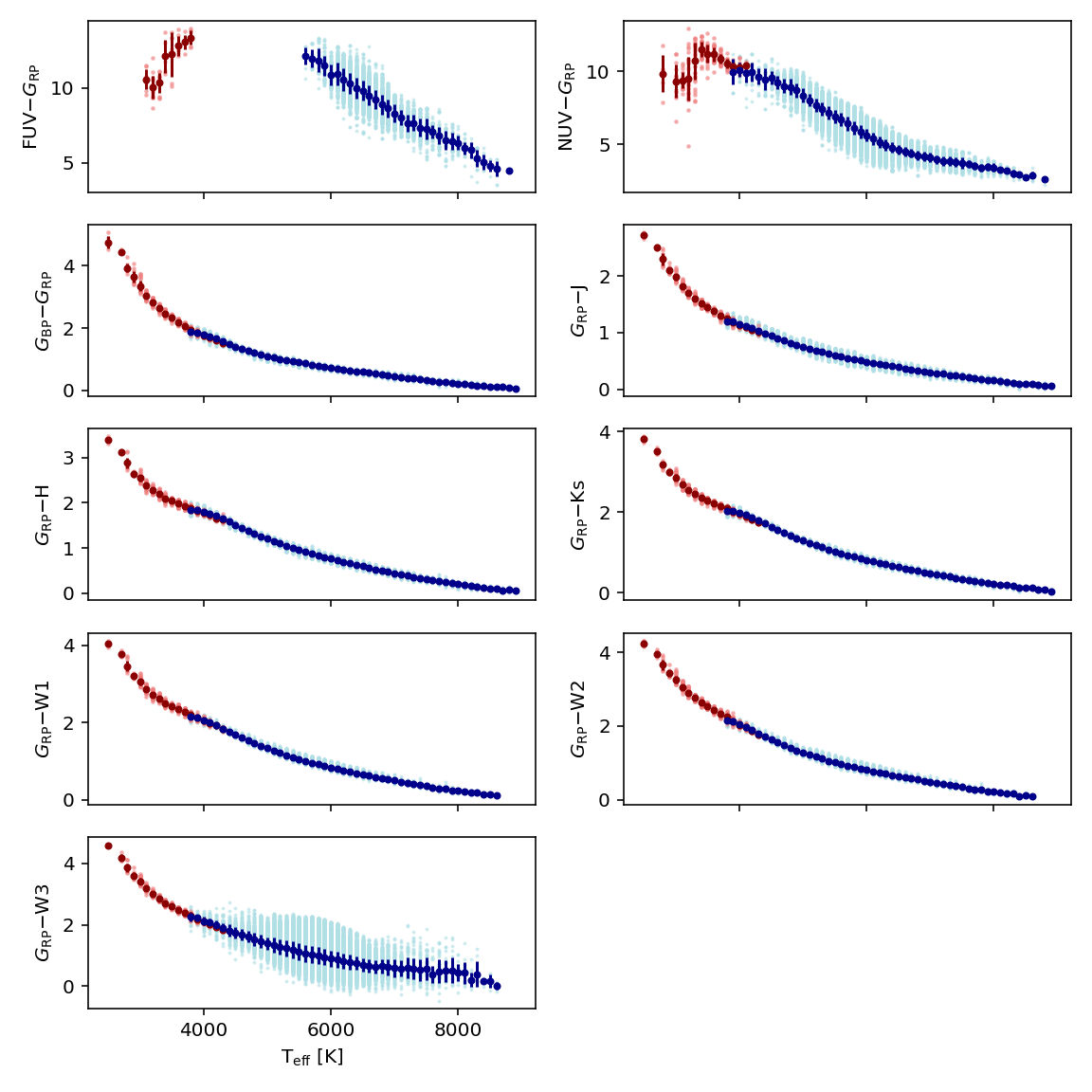}
    \caption{Empirical colour -- effective temperature (T$_{\rm eff}$) relations late-type stars. The mean and sample standard deviation in each T$_{\rm eff}$ bin are shown as a solid point with an error bar. These have been calculated from the points in the same T$_{\rm eff}$ bin plotted using dots.}
    \label{fig:color-teff}
\end{figure*}

The empirical relations derived are shown in Fig.~\ref{fig:color-teff}.
The tabulated colour\,--\,T$_{\rm eff}$ relations for the CARMENES and GALAH have been stored separately and the user has the option to choose which relation to use in the range from 3800\,K to 4700\,K where these tabulations overlap.

\bibliographystyle{mnras}
\bibliography{allbib} 


\bsp	
\label{lastpage}
\end{document}